# Pathobiological Dictionary Defining Pathomics and Texture Features: Addressing Understandable AI Issues in Personalized Liver Cancer; Dictionary Version LCP1.0

Mohammad R. Salmanpour[1,2,3*†], Seyed Mohammad Piri[4†], Somayeh Sadat Mehrnia[5], Ahmad Shariftabrizi[6], Masume Allahmoradi[7], Venkata SK. Manem[8,9], Arman Rahmim[1,2], Ilker Hacihaliloglu[1,10]

[1]Department of Radiology, University of British Columbia, Vancouver, BC, Canada
[2]Department of Integrative Oncology, BC Cancer Research Institute, Vancouver, BC, Canada
[3]Technological Virtual Collaboration (TECVICO Corp.), Vancouver, BC, Canada
[4]Cancer Institute of Iran, Tehran University of Medical Sciences, Tehran, Iran
[5]Department of Integrative Oncology, Breast Cancer Research Center, Motamed Cancer Institute, ACECR, Tehran, Iran
[6]Department of Radiology, University of Iowa Carver College of Medicine, IA, US
[7]Department of Pathology, Iran University of Medical Sciences, Tehran, Iran
[8]Centre de Recherche du CHU de Québec, Université Laval, QC, Canada
[9]Département de Biologie Moléculaire, Biochimie Médicale et Pathologie, Université Laval, QC, Canada
[10]Department of Medicine, University of British Columbia, Vancouver, BC, Canada

([†]) Co-First Authors: These authors had equal contributions as co-first authors.

(*) Corresponding Author: Mohammad R. Salmanpour, PhD; *msalman@bccrc.ca*

**ABSTRACT**

**Background:** Artificial intelligence (AI) holds strong potential for medical diagnostics, yet its clinical adoption is limited by a lack of interpretability and generalizability. This study introduces the Pathobiological Dictionary for Liver Cancer (LCP1.0), a practical framework designed to translate complex Pathomics and Radiomics Features (PF/RF) into clinically meaningful insights aligned with existing diagnostic workflows.
**Methods:** QuPath and PyRadiomics, standardized according to IBSI guidelines, were used to extract 333 imaging features from hepatocellular carcinoma (HCC) tissue samples, including 240 PF-based-cell detection/intensity, 74 RF-based texture, and 19 RF-based first-order features. Expert-defined 1000×1000 μm ROIs from the public dataset excluded artifact-prone areas, and features were aggregated at the case level. Their relevance to the WHO grading system was assessed using multiple classifiers linked with feature selectors. The resulting dictionary was validated by 8 experts in oncology&pathology.
**Results:** In collaboration with 10 domain experts, we developed a Pathobiological dictionary of imaging features such as PFs and RF. In our study, the Variable Threshold feature selection algorithm combined with the SVM model achieved the highest accuracy (0.80±0.01, $P<0.05$), selecting 20 key features—primarily clinical and pathomics traits such as Centroid, Cell Nucleus, and Cytoplasmic characteristics. These features, particularly nuclear and cytoplasmic, were strongly associated with tumor grading and prognosis, reflecting atypia indicators like pleomorphism, hyperchromasia, and cellular orientation.
**Conclusion:** The LCP1.0 provides a clinically validated bridge between AI outputs and expert interpretation, enhancing model transparency and usability. Aligning AI-derived features with clinical semantics supports the development of interpretable, trustworthy diagnostic tools for liver cancer pathology.

## 1. INTRODUCTION

Hepatocellular carcinoma (HCC) is the most prevalent form of primary liver cancer, accounting for nearly 90% of all cases. It is the sixth most common cancer globally and the third leading cause of cancer-related mortality [1]. Despite advancements in treatment, HCC remains a major healthcare burden, with an overall survival rate of less than 8% [2]. Its incidence and mortality rate is projected to rise by over 55% by 2040, leading to an estimated 1.3 million deaths worldwide, with the highest prevalence observed in Asia [3, 1], but its incidence and mortality rates are increasing rapidly in the United States and Europe [4]. The development of HCC is influenced by various risk factors, including, alcohol consumption [5] chronic infections such as hepatitis B and C, metabolic disorders, cirrhosis [6, 7], exposure to carcinogens like aflatoxin, obesity, diabetes [8], genetic mutations, and epigenetic changes [9]. Over time, the role of viral infections has declined, while metabolic-associated fatty liver disease (MAFLD) has become a more prominent contributor to HCC incidence [10].

The heterogeneity of HCC poses significant challenges in diagnosis, prognosis, and treatment planning. Traditional histopathological analysis remains the gold standard for tumor characterization, yet it often lacks the ability to capture the full complexity of tumor biology. In recent years, computational pathology has emerged as a promising approach to enhance cancer diagnostics. Pathomics Features (PFs), the extraction and analysis of quantitative features from histopathological images, enables deeper insights into tumor morphology, heterogeneity, and progression. By leveraging ex vivo imaging (known as histomics), pathomics offers a data-driven method to complement traditional pathology. Studies in this field have demonstrated its potential to bridge the gap between histopathology and computational analysis [11, 12, 13]. However, challenges remain in ensuring the interpretability and pathological/biological relevance of extracted features [14, 15, 16].

A critical limitation in current pathomics-based research is the lack of transparency in Artificial Intelligence (AI)-driven models. Many AI-based diagnostic tools function as "black boxes," making it difficult for clinicians to understand the rationale behind predictions [17, 18, 19]. While post hoc explainability techniques attempt to address this issue, they often fail to establish a direct pathological/biological connection between extracted quantitative and semantic features. There is an increasing need to integrate biologically meaningful analyses during model development to ensure that AI-generated insights align with established pathological and clinical knowledge [20, 21]. The lack of clear pathological and biological definitions in multi-omics studies further complicates the extraction of robust conclusions. [17].

Efforts to integrate Radiomics Features (RFs), PFs, and genomics have been made to enhance predictive models for both clinical [22] and biological [23, 24, 25, 26] outcomes. However, unsupervised clustering of high-dimensional imaging often fails to reveal biologically relevant patterns [17]. Similarly, studies in radio-pathomics, which aim to develop virtual biopsies by combining radiological and histopathological data, face challenges in ensuring the biological interpretability of combined models [27, 28, 29, 30, 31]. Even when these models demonstrate strong statistical performance, they often lack clear biological explanations for their predictions [32, 33]. Tomaszewski [17] emphasized that future studies must prioritize biologically meaningful analyses in model development and validation to generate hypotheses about the underlying mechanisms driving observed relationships.

A major issue in many pathomics studies is the reliance on statistical correlations without an explicit connection to biological mechanisms [34, 35, 17, 30]. Furthermore, computationally derived features often lack direct correspondence with visual semantic characteristics that pathologists use in clinical assessments. Addressing this gap requires a standardized framework, such as a structured dictionary or index, to map computational features to clinically and pathologically interpretable descriptors. Such a framework would enhance model transparency, improve reproducibility, and facilitate collaboration between computational researchers and medical professionals [36].

The integration of multi-omics data, including PFs, RFs [37, 38, 39], and genomics [40, 41, 42], has shown great potential in advancing cancer diagnostics and prognostics [43]. By linking microscopic histological structures with imaging-based patterns and molecular markers, multi-omics approaches provide a more comprehensive understanding of tumor behavior, heterogeneity, and patient-specific prognosis. However, effective interpretation of these features requires a structured pathomics dictionary or index to unify terminology, ensure standardization, and facilitate cross-study comparisons [44]. This dictionary would help identify surrogate biomarkers, establish correspondences across datasets, and improve interdisciplinary communication [45].

Furthermore, understanding deep learning-extracted features requires a clear distinction between first-order handcrafted features and more complex, learned representations [46, 47, 48, 49]. A well-defined pathomics dictionary would enhance repeatability and reproducibility in computational pathology research [50, 51]. Many studies face challenges due to limited datasets [17, 52], variations in feature extraction methodologies [53, 54], and inconsistencies in measurement techniques [55, 56, 57], reinforcing the need for a standardized feature definition approach. Developing a comprehensive pathomics dictionary would not only enhance reproducibility but also improve the interpretability of AI-driven models, ultimately making them more applicable in clinical practice.

To address these challenges, this study proposes a structured framework to map PFs to clinically interpretable characteristics. First, we utilized visual semantic feature definitions based on the World Health Organization (WHO) grading system to establish conceptual relationships between imaging-derived PFs and visually recognizable histopathological patterns. Next, we apply a range of compendium of Machine Learning (ML) techniques, including interpretable feature selection algorithms (FSA) combined with a mix of interpretable and complex classifiers, to predict HCC overall survival outcomes. Finally, we demonstrate how the proposed pathomics dictionary enhances feature interpretability, improves model performance, and validates the clinical relevance of the selected features. By aligning computationally extracted PFs with established pathological criteria, this study aims to enhance transparency and biological interpretability in AI-driven cancer research. Our approach has the potential to advance precision oncology by fostering collaboration between computational scientists and clinicians, ultimately improving patient outcomes and optimizing treatment strategies [58].

## 2. MATERIAL AND METHOD
### 2.1. Exploring the Relationship Between PFs and the WHO Grading System

This study utilized QuPath (https://qupath.github.io/) and PyRadiomics (pyradiomics.readthedocs.io, standardized according to the IBSI [59]), to extract 333 quantitative PFs and RFs from HCC tissue samples. These features, detailed in Supplemental Tables S1, S2, and sections 4 and 5 of the supplemental file, included 240 PF-based cell detection and color intensity measures, 74 RF-based texture features (23 Gray Level Co-occurrence Matrix Features (GLCM), 16 Gray Level Size Zone Matrix Features (GLSZM), 16 Gray Level Run Length Matrix Features (GLRLM), 5 Neighboring Gray Tone Difference Matrix Features (NGTDM), and 14 Gray Level Dependence Matrix Features (GLDM)), and 19 RF-based first-order features. These features were analyzed to assess their semantic relevance to the WHO grading system (4$^{th}$ and 5$^{th}$ edition) for HCC (Supplemental Tables S3 and S4; Figure 1), which classifies tumors based on histological differentiation and serves as a key prognostic tool. While the classic Edmondson and Steiner grading system [60] is similar and still used in practice, the updated 3-tiered WHO classification was adopted as the standard, though we also considered relevant aspects of the classic grading system. We also used the previous WHO update in some places to make the dictionary more comprehensive. The 1000 x 1000-pixel Region of Interests (ROIs), selected by an experienced pathologist, excluded areas with artifacts (e.g., excessive fat droplets, clear cells, or overlapping nuclei) due to challenges in accurately calculating cell layer numbers in such regions. Color normalization and feature measurements were aggregated at the case level using mean values, and edge effects were assessed with four smoothness levels (50, 100, 150, and 200 micrometers, μm). Nuclear segmentation and PFs extraction were performed using QuPath's Groovy scripts, with the exported data statistically analyzed in Excel. The study focused exclusively on the relationship between extracted features and the WHO grading system, excluding factors such as histologic variants or Tumor, Nodes, and Metastasis (TNM) staging due to data heterogeneity. QuPath was chosen for its precise cell segmentation, ease of use, and comprehensive feature set, though other software (such as Fiji and Cell Profiler) could yield similar results. Furthermore, this dictionary was validated by three experienced pathologists, an experienced physician, an experienced biologist, and three experienced medical physicists who are familiar with PFs, RFs, and AI analysis. Because the details of the PFs and RFs, as well as the semantic features, are extensive, we have provided them in the supplemental file for further reference.

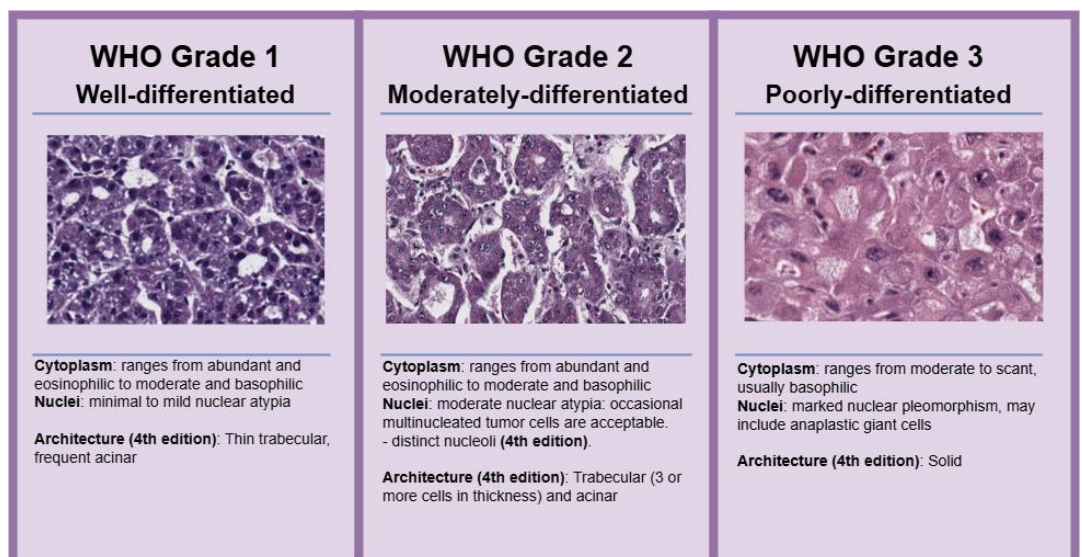

**Fig. 1.** Detailed figure summarizing the World Health Organization (WHO) Grading System for Hepatocellular Carcinoma (HCC); 4[th] and 5[th] edition, incorporating the essential characteristics for each grade. This image showcases three microscopic views of HCC tissue samples presented in updated version of WHO grading system (5[th]; 2019),Each panel represents a different grade (1-3 represents well, moderately, and poorly differentiated respectively), illustrating the progressive changes in cytoplasm, nuclei, and cellular architecture features. Higher grades reflect increased atypia and a poorer prognosis. Note that in the WHO grading update, architectural features are not mentioned in the criteria and are only mentioned in the description of the grading types. However, in the previous edition (4[th]), architecture was mentioned as part of the criteria. To access deeper and more features and also to access a more complete definition of the semantic relationship of the features, we considered the two criteria. With more attention to the WHO 2019 criteria, but we have considered more definitions in a way that we have a greater coverage of the semantic features related to the grading of hepatocellular carcinoma.

## 2.2. Patient Data

We utilized a publicly available dataset of liver cancer patients (217 subjects) from the Liver Hepatocellular Carcinoma (LIHC) project [61]. This international initiative, part of The Cancer Genome Atlas (TCGA), aims to comprehensively characterize liver cancer (specifically HCC) genomic and molecular levels to better understand its causes. The LIHC project, which includes data from the Cancer Imaging Archive (TCIA) and the GDC Data Portal [62, 63], seeks to identify genetic and molecular alterations driving the disease to improve targeted therapies. The dataset includes whole slide images of HCC tumor tissue linked to different pathological data and the overall survival outcome of the patient. We excluded images related to cholangiocarcinoma and healthy individuals, which were a small part of the dataset. We extracted pathology slide data from 217 HCC patients from TCIA [61], residing in the Genomic Data Commons (GDC) Data Portal [64] using QuPath and PyRadiomics software, and at the same time, we evaluated the predictive power of the model for predicting overall survival in HCC patients (Figure 2). In this study, we collected and analyzed demographic and clinical data, including variables such as age, race, gender, prior treatment before surgery, treatment after surgery, treatment type, AJCC pathologic stage (TNM), prior malignancy, year of diagnosis, and tumor grade to understand their impact on patient outcomes and disease progression. The samples were composed of 60.43% men and 39.57% women, with the average age of the patients estimated to be around 33.89 years.

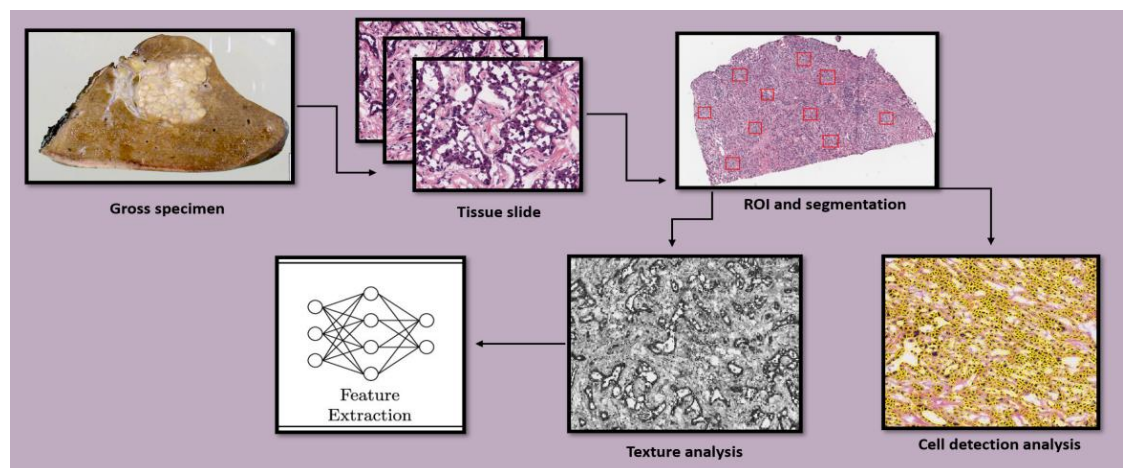

**Fig. 2.** Workflow figure outlines the steps from specimen resection, and ROI (Region of Interest) selection to obtaining RGB (Red, Green, Blue) and gray-scale images, enabling the extraction of relevant features. After selecting a suitable ROI, the images are converted into two types, RGB and grayscale, for mask creation and feature extraction. The data extraction process is performed using the cell detection modules of the QuPath software and radiomics modules from PyRadiomics.

## 2.3. Understandable Classification Methods

To select relevant features for classification, nine interpretable FSAs – Chi-Square Test (CST), Correlation Coefficient (CC), Mutual Information (MIS), Variance Threshold (VTS), ANOVA F-test (AFT), Information Gain (IGS), Univariate Feature Selection (UFS), Fisher Score (FSF) [65, 66], and Least Absolute Shrinkage and Selection Operator (LASSO) [67, 68] (detailed in Supplemental Section 1.2.2.1) – were applied to the training set (see Supplemental Section 1.2.1). The selected features were then used with thirteen classification algorithms (CAs), including six interpretable models, Decision Tree Classifier (DTC) [69], Logistic Regression Classifier (LOC) [70], Linear Discriminant Analysis (LDA) [71], Naïve Bayes Classifier (NBC), k-Nearest Neighbors (KNN), Rule-Based Classifier (RUC) [72] and seven black-box ensembles, Random Forest Classifier (RFC) [73], Extreme Gradient Boosting (XGB), Light Gradient Boosting Machine (LGB), Cat Boost Classifier (CBC) [74], Support Vector Machine (SVM) [38], Stacking Classifier (STC), and Multi-Layer Perceptron (MLP) [75] (described in Supplemental Section 1.2.2.2). Interpretable ML, also known as Interpretable AI, focuses on creating transparent models [76]. As ML systems increasingly impact societal decisions, the "black box" nature of many models, which obscures their prediction logic, becomes problematic [69]. This lack of transparency is especially critical in regulated sectors where auditability is essential. Consequently, there's a growing need for interpretable ML algorithms that offer both accurate predictions and understandable decision-making processes [77, 78, 79]. This section explores the significance of these algorithms, methods for enhancing interpretability, and the challenges in evaluating explanation quality [80, 81, 82, 83]. The proposed workflow involves the following steps (Fig. 3):

  i. Identifying cells in tissue images and gathering quantitative data about their morphology, staining, and spatial relationships in QuPath.
  ii. Expert pathologistsdelineated cancerous lesion masks on whole slide images (WSIs) to accurately segment Regions Of Interest (ROI).
  iii. QuPath extracts biological, structural, and spatial features at the cellular/tissue level, whereas PyRadiomics focuses on computationally advanced features, especially for texture and intensity analysis, often useful in building predictive models. Together, they can efficiently provide robust and multi-dimensional data from digital pathology and medical images.
  iv. Patient datasets were separated into two parts: 85% of the data was allocated for five-fold cross-validation, while the remaining 15% served as an external nested testing set to ensure a robust evaluation of the model's generalizability.
  v. The extracted quantitative features were scaled using the min-max normalization, applied based on the parameters of the training set.

vi. Nine interpretable FSAs were utilized on the training datasets to identify the most relevant feature subsets for classification tasks. These included CST, CC, MIS, VTS, AFT, IGS, UFS, FSF, and LASSO.

vii. A total of 13 CAs were applied to the selected features, consisting of six interpretable models and seven black-box ensemble techniques. The interpretable models included DTC, LOC, LDA, NBC, KNN, and RUC. The black-box ensemble techniques comprised RFC, XGB, LGB, CBC, SVM, STC, and MLP.

viii. The top 20 features identified by the FSAs, based on their classification performance, were chosen and organized into a dictionary (Table 1) to enhance interpretability and clinical relevance associated with the selected imaging features [84]. Details of all algorithms are described in supplemental section 1.2. Incorporating this dictionary improves interpretability and explainability, enabling clinicians to understand model decisions and the impact of features on liver cancer classification. For AI developers, it offers clear insights into feature importance, supporting transparent model design, enhanced performance, and better alignment with clinical needs.

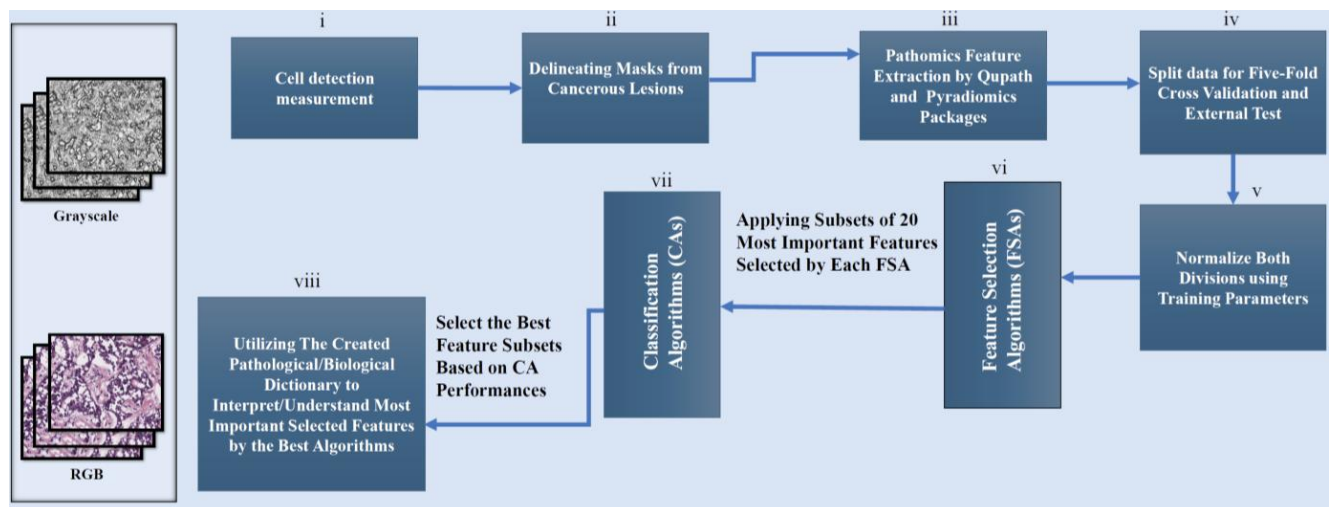

**Fig. 3** Workflow for cancer classification using a combination of cell detection measurements from whole slide images (WSIs) and subsequent feature extraction and selection steps. The process involves i) cell detection measurement, ii) defining and marking the boundaries of cancerous areas, iii) feature extraction using QuPath and PyRadiomics, iv) data splitting for cross-validation and external nested testing, v) Applying normalization using parameters learned from the training set, vi) Feature selection was performed on the training sets of different datasets using a suite of nine feature selection algorithms (FSAs), vii) A diverse set of 13 classification algorithms (CAs), ranging from interpretable models to more complex methods, were applied to the feature subsets identified by the (FSAs, and, viii) interpretation of the most (20) important selected features using a created pathological/biological dictionary. The diagram highlights the integration of image analysis and Machine earning (ML) for improved diagnostic accuracy.

## 3. RESULTS
### 3.1. Creation of Pathological/Biological Dictionary

This study investigated the relationship between PFs and tumor cell characteristics (specifically cytologic features as a main diagnostic criteria and architectural pattern as an ancillary morphologic feature) in HCC, using visual semantic features from the 5$^{th}$ edition of WHO grading system [85] (See Supplemental File: Table S5). While our analysis focused on the Liver cancer dataset, the resulting common dictionary of PFs has broader applicability to other cancers. HCC grading is a crucial prognostic factor [86], traditionally categorized into four levels based on Edmondson and Steiner's 1954 classification [60]. The simplified three-tier WHO system; 5$^{th}$ edition (well, moderately, poorly differentiated) aligns with current practice, with the worst grade dictating the overall prognosis. This reflects the significant prognostic impact of distinguishing between well and poorly differentiated HCCs; Well-differentiated HCCs closely mimic normal hepatocytes, while Poorly differentiated HCCs lack hepatocellular characteristics. Moderately differentiated HCCs exhibit malignant features, with morphology indicative of hepatocellular differentiation; moderate nuclear atypia is present.. In summary, descriptions of certain features are provided for one grade and can be generalized across grades (Supplemental File; Table S5).

### 3.1.1. Grade 1 WHO Classification (Well Differentiated)

Grade 1 HCC typically offers a better prognosis, characterized by slower tumor growth and greater treatment responsiveness, leading to prolonged symptom-free periods and improved recovery chances. The visual semantic features associated with Grade 1 (WHO Classification) are quantifiable using various PFs. Grade 1 tumor cells resemble mature hepatocytes with minimal atypia, mainly exhibiting thin trabecular and acinar architectures and cytoplasmic staining ranging from abundant eosinophilic to moderate basophilic. Minimal nuclear atypia is consistent with fatty change, a common morphological feature of early HCC [87] (See Table S3).

**3.1.1.1 Cytologic visual semantic features**

Cytological features, visualized using Hematoxylin and Eosin (H&E) staining, are quantified by PFs such as H&E Optical Density (OD) features from the cell detection module. Basophilic staining (higher hematoxylin absorption) indicates transcriptional and translational activity, while eosinophilic staining (higher eosin absorption) suggests increased mitochondrial activity and protein synthesis. The nucleus-to-cytoplasm (N/C) ratio, measured in square micrometers, reflects cytoplasmic abundance. Nuclear atypia, assessed by PFs such as nuclear area, perimeter, circularity, caliper dimensions, and eccentricity, increases with the larger nuclear area, perimeter, and max caliper, and higher eccentricity, along with lower circularity and min caliper. Visual semantic features and the associated PFs/RFs serve as tools to evaluate WHO Grade 1 lesions, as follows:

- Cytoplasm Hematoxylin Optical Density (CHOD) represents the intensity of basophilic dye absorption in the cytoplasm to show basophilia which is typically an indicator of the presence and distribution of RNA and acidic proteins in the cell's cytoplasm. High hematoxylin OD in the cytoplasm may signify active protein synthesis, as ribosomes and RNA are more abundant in metabolically active cells.
- Cytoplasm Eosin Optical Density (CEOD) represents the extent to which the cytoplasm absorbs eosin dye. Abnormal eosin OD may signify changes in protein content or cytoplasmic density, often associated with pathological conditions like inflammation or neoplasia. A high eosin OD indicates a higher concentration of basic proteins in the cytoplasm.
- Nucleus Cell Area Ratio (NCAR) represents the proportion of the total cell area occupied by the nucleus. This feature describes the relative size of the nucleus compared to the overall cell size. Cells with a high nucleus-to-cell area ratio often have increased nuclear material (e.g., chromatin or DNA content), which may indicate higher transcriptional activity and cellular growth. A higher ratio is often seen in malignant or dysplastic cells, where nuclear enlargement occurs due to increased DNA content or abnormal nuclear morphology (e.g., small cell change). A lower ratio may be associated with conditions where cytoplasmic volume increases relative to the nucleus (e.g., large cell change). Changes in the nucleus-to-cytoplasm ratio also aid in identifying nuclear pleomorphism, which is one of the semantic features associated with Grade 3.
- Nucleus Area (NA) represents the physical size or surface area of the cell's nucleus and reflects the overall size of the nucleus. Larger nuclei are often associated with increased DNA content (aneuploidy) or active transcription in cancer cells. Variability in the size of nuclei (including nucleus area) across a tissue sample can signify abnormal cellular differentiation, common in higher-grade malignancies.
- Nucleus Perimeter (NP) reflects the overall size of the nucleus, complementing measurements like the nucleus area. A highly irregular or elongated nuclear perimeter is often indicative of nuclear pleomorphism, a hallmark of dysplastic or malignant cells (usually initiated from Grade 3).
- Nucleus Circularity (NC) represents the shape of the nucleus; specifically, how close shape is to a perfect circle. A perfect circle has a circularity value of 1 and Values closer to 0 indicate irregular or elongated nuclear shapes. Although this cannot be generalized in general, circular nuclei are generally associated with benign or normal cells. On the other hand, Variations in nuclear circularity within a tissue sample are indicative of heterogeneity, a common sign of higher-grade tumors or aggressive malignancies Which is mostly seen in intermediate differentiation onwards. cylindrical or cigar-shaped formation was a sign of atypia

- Nucleus Max and Min Caliper refer to measurements of the largest and smallest diameters of the nucleus, effectively describing its dimensions and shape. The max and min calipers provide information on the nuclear orientation, helping to study cellular arrangement and alignment within tissues. This feature is important in both cytologic semantic features and architectural pattern where the structure is disrupted in the aggressive grades, the polarity of the cells is lost, and they move towards solid architecture (Grade 3).
- Nucleus Eccentricity represents the degree to which the shape of the nucleus deviates from being perfectly circular or symmetrical. Increased Eccentricity indicates irregular or elongated nuclei, which are commonly observed in dysplasia. It helps quantify the extent of nuclear distortion, which is useful in identifying neoplastic changes

### 3.1.1.2 Architectural Visual Semantic Features

Architecturally, Grade 1 tumors resemble normal tissue, requiring differentiation from dysplastic nodules and adenomas. Dysplastic nodules, found in cirrhotic livers, have sharper reticulin frameworks and less pronounced nuclear atypia than Grade 1 HCC. Reticulin staining is crucial for assessing reticulin network loss. The thin trabecular architecture of Grade 1 is assessed using PFs like GLCM contrast (GLCM_Co), correlation (GLCM_Corr), joint energy (GLCM_JE), and inverse difference moment (GLCM_IDM). Correlation reflects the linear dependency between gray levels of neighboring pixels (trabecular vs. solid pattern), while joint energy quantifies the frequency of gray level pairs. Delaunay features (mean, median, max, min) from the cell detection module quantify spatial regularity and distribution uniformity, identifying architectural changes and potentially indicating disease progression.

PFs/RFs serve as tools to evaluate WHO Grade 1 lesions, as follows:
- Contrast from the GLCM category (GLCM_Co) represents the variability in intensity within the cellular texture, particularly reflecting the heterogeneity of the cell structure or surrounding tissue texture. Tumors often display higher textural contrast due to heterogeneous cellular arrangements, abnormal chromatin distribution, or variations in staining uptake. Higher contrast values often indicate regions with greater structural complexity or disruption, such as nuclear irregularities or cytoplasmic granularity commonly seen in dysplastic or malignant cells. Architecturally, cells with disordered organization tend to exhibit higher contrast, and in HCC, the contrast increases with higher tumor grade due to increased tissue heterogeneity. However, exceptions exist—solid architectural patterns may show lower microenvironmental contrast despite significant intracellular heterogeneity. Therefore, assessing contrast across various areas of the tumor is recommended for a more comprehensive evaluation.
- Correlation from the GLCM category (GLCM_Corr) is a measurement ranging from 0 (no correlation) to 1 (perfect correlation), reflecting the linear relationship between gray level intensity values and their corresponding voxel positions in the GLCM. The correlation feature assesses the linear relationship between the gray levels of neighboring pixels. High correlation values indicate strong linear dependency between pixel intensities, often associated with more uniform textures and lower heterogeneity. Conversely, low correlation values reflect weaker dependency, signifying greater variability and higher heterogeneity in the texture.
- Joint Energy from the GLCM category (GLCM_JE) assesses the randomness or variation in the distribution of neighboring intensity values, capturing the texture's heterogeneity. Joint energy measures the prevalence of homogeneous patterns within an image. Higher values indicate more frequent occurrences of specific gray level pairs, reflecting a more uniform and consistent texture. This is particularly significant for analyzing intra-nuclear texture features and determining architectural characteristics.
- Inverse Difference Moment from the GLCM category (GLCM_IDM) evaluates the local homogeneity of an image. IDM (Inverse Difference Moment) assigns weights inversely proportional to contrast weights, decreasing exponentially as values move away from the diagonal (i=j) in the GLCM. It reflects the similarity of pixel intensities within a region, where higher IDM values indicate low variability, high local similarity, and greater homogeneity and uniformity.

**3.1.2. Grade 2 WHO Classification (Moderately Differentiated)**

Grade 2 HCC represents an intermediate stage between well and poorly differentiated HCC, carrying an intermediate prognosis. Early diagnosis, thorough histopathological examination, and digital pathology tools for feature quantification are crucial. Careful assessment of architectural disorganization, nuclear atypia, and reticulin loss helps distinguish Grade 2 from well and poorly-differentiated lesions, improving treatment planning and outcomes. Grade 2 HCC cells typically exhibit basophilic rather than eosinophilic cytoplasm, hyperchromatic nuclei with prominent nucleoli, and sometimes multinucleated giant cells. Unlike the thin trabecular pattern of well-differentiated HCC, Grade 2 shows thicker trabeculae (3 or more cells thick) reflecting more hepatocyte de-differentiation. Reticulin or CD34 staining aids in identifying trabecular patterns. Since cellular differentiation is less in grade 2 than in grade 1, the amount of pseudo-acinar structures increases and the trabecular layers become thicker, textural PFs—GLCM_Co, GLCM_Corr, GLCM_JE, and GLCM_IDM become more conclusive in representation of the architecture. Higher contrast indicates greater intensity variation, correlation reflects linear dependency of pixel intensities (lower values indicating heterogeneity), joint energy quantifies homogeneous patterns (higher values indicating uniformity), and inverse difference moment (IDM) assesses pixel intensity similarity (higher values indicating homogeneity). PFs (cell detection module); nuclear area, perimeter, circularity, caliper dimensions, and eccentricity reflect nuclear pleomorphism, atypia, N/C, irregular nuclear membranes, multinucleation, and prominent nucleoli. Nuclear hyperchromasia is represented by mean nuclear H&E OD features (mean, sum, standard deviation, max, min, range). Delaunay features analyze the spatial arrangement of nuclei, providing insights into trabecular architecture and cellular organization [88]. In Grade 2, moderately thickened trabeculae (2-3 cell layers), correlate with the partial loss of reticulin fibers [89].

As mentioned in the text, there are many similarities between Grade 1 and Grade 2 (Well-to-Moderate differentiation). The features described for Grade 1 are also applicable to Grade 2 (Cytoplasm Hematoxylin OD, Cytoplasm Eosin OD, Nucleus Cell Area Ratio, Nucleus Area, Nucleus Perimeter, Nucleus Circularity, Nucleus Eccentricity, and Nucleus Max and Min Caliper). The notable difference is that in Grade 2, semantic features indicate greater aggressiveness compared to Grade 1. One of the key factors in distinguishing these two grades is the presence of moderate nuclear atypia with occasional multinucleated tumor cells which can be assessed by Nucleus Cell Area Ratio, Nucleus Area, Nucleus Perimeter, Nucleus Circularity, Nucleus Eccentricity, and Nucleus Max and Min Caliper and distinct nucleoli assessed by GLCM_contrast and GLCM_entropy. According to the WHO grading system, the architecture is trabecular, with 3 or more cells in each layer. From this grade onward, the structure tends to thicken and progress until it eventually transforms into a solid structure. Note that the distinction between grades 1 and 2 is sometimes very difficult, and noticeable morphological changes are mainly noticeable in grade 3.

### 3.1.3. Grade 3 WHO Classification (Poorly differentiated)

Advanced HCC exhibits expansive, infiltrative growth with complete neovascularization, often including unpaired arteries and vascular infiltration. Portal tracts are absent within the tumor, and all classical architectural patterns (trabecular/sinusoidal, pseudo-glandular, and solid) are typically present [57]. Given the similarities between Grades 1 and 2, and the marked differences starting at Grade 3, Grades 1 and 2 can be grouped as well-to-moderately differentiated, and Grade 3 as poorly differentiated HCC. Detailed texture feature analysis is primarily relevant for Grades 3 and above. Grade 3 (poorly differentiated) HCC differ significantly from normal tissue, exhibiting a poor prognosis and increased mortality risk. Marked cellular atypia and disorganized tissue structure are characteristic, typically presenting as a solid architecture with cells in nests, lacking space between them. The cytoplasm is moderate to scant, usually basophilic; nuclei show marked pleomorphism and may include anaplastic giant cells. Quantitative analysis of these features could potentially predict disease progression and outcomes based on tumor heterogeneity.

#### 3.1.3.1 Cytologic Visual Semantic Features

- Cytoplasm Hematoxylin Optical Density (CHOD) at higher grades usually becomes more basophilic, and this has already been explained. A high mean value would suggest that the tissue sample is heavily stained with Hematoxylin, indicating a higher concentration of ribosome in the cell.
- Nucleus Cell Area Ratio (NCAR) represents a high value because of the presence of cells with scant cytoplasm and large nuclear area. This feature could be related to anaplastic giant cells' semantic feature.
- Nucleus size, Roundness, and GLCM Contrast could show marked nuclear pleomorphism. A higher GLCM contrast value means higher heterogeneity in both cell features and architecture and could be related to intranuclear heterogeneity, the presence of nucleoli, and nuclear atypia. This is the feature of anaplastic giant cells also.

10Salmanpour et al.     Dictionary Version LCP1.0

- MEAN Cell Area (MCA) is the average of the cell areas calculated from multiple cells expressed in square micrometers (µm²) or pixels within the region of interest. Enlarged cell areas may result from cellular atypia, pleomorphism, or increased cytoplasmic volume, often associated with malignancy. Variations in cell areas across different regions in the tumor can help identify heterogeneity, which is a hallmark of aggressive tumors.
- MEAN Cell Perimeter (MCP) is the average length of the boundary of the cell. Irregular or elongated cell perimeters often correlate with dysplasia or malignant transformation.
- Increased variability in the cell perimeter may signify invasive behavior in the tumor, as cells lose their structural uniformity in higher-grade HCC.
- MEAN Cell Circularity (MCC) is a measure of how closely the shape of a cell resembles a perfect circle. Cells in normal liver tissue tend to have a regular circular shape, while dysplastic or malignant cells are often irregular, elongated, or distorted. Greater deviations from circularity may be correlated with the tumor grade, invasion, or loss of cell polarity.
- MEAN Cell MinCaliper (MCMxC) and MaxCaliper (MCMnC) are the smallest and greatest possible measurement of a cell. High-grade HCC cells often show increased variation in their minimum and maximum caliper values, representing a mixture of abnormal cell enlargement and shrinking.
- MEAN Cell Eccentricity (MCE) measures how much a shape deviates from being circular. Higher eccentricity values may also correspond to disrupted architecture and heterogeneity in higher HCC grades.

**3.1.3.2 Architectural visual semantic features**
In Poorly Differentiated HCC, Sinusoid-like blood spaces are absent, and the presence of vascular structures without venous components or bile ducts reflects disorganized tissue structure. The increased contrast between tumoral and normal tissue, along with increased intra-tumoral heterogeneity in solid architectures, makes texture features highly effective in this grade. A wide range of PFs, including homogeneity, heterogeneity, uniformity, and intensity-related features, are relevant (see Supplemental Table S5).

    RFs (first-order statistics, shape properties, GLCM, GLRLM, GLSZM, GLDM, and NGTDM) are valuable tools for assessing Grades 3 and 4 HCC, quantifying tumor architecture, heterogeneity, and intensity distribution. These features enable precise grading and prognosis prediction, supporting non-invasive classification and clinical decision-making via advanced ML. These features differentiate low-grade (Grades 1 and 2) from high-grade (Grades 3) HCC (Edmondson-Steiner or WHO classification). First-order features (FOs) in high-grade tumors (Grades 3) show increased heterogeneity (higher entropy), reflecting their aggressive nature. Mean intensity is lower in necrotic areas. High-grade tumors have irregular shapes (increased surface area and perimeter), decreased sphericity and compactness indicating invasive growth. GLCM metrics analyze texture heterogeneity, with high-grade HCC exhibiting increased contrast and decreased homogeneity. GLSZM features measure the size of homogeneous gray-level zones; high-grade tumors show larger zones (increased large zone emphasis). GLRLM features quantify lengths of consecutive pixels with similar intensities; high-grade tumors could show stronger long run emphasis due to necrosis. NGTDM features capture intensity differences between a pixel and its neighbors; high-grade HCC shows increased busyness, complexity, and contrast. Finally, GLDM features quantify gray-level dependence; high-grade tumors exhibit increased dependence entropy and large dependence emphasis due to their irregular texture.

- Contrast from the GLCM category (GLCM_Co) was explained in previous sections. Cells that lack organized structure typically show enhanced contrast. In the case of HCC, more aggressive (higher-grade) tumors display greater contrast due to their more varied tissue composition. Yet, this isn't a universal rule—some tumors with solid growth patterns might show reduced contrast between different areas, even when the cells themselves are quite diverse internally. As a result, it's best practice to examine contrast in multiple regions of the tumor to get a complete understanding of its characteristics.
- Inverse Difference Normalized from the GLCM category (GLCM_IDN) is another measure of the local homogeneity of an image. IDN normalizes the difference between the neighbouring intensity values by dividing over the total number of discrete intensity values. GLCM_IDN primarily represents tissue uniformity or homogeneity and is inversely linked to texture variation or heterogeneity within a tissue sample. Higher GLCM_IDN values indicate smoother transitions between adjacent gray levels, reflecting more consistent and less varied tissue patterns. In contrast, lower values point to greater structural variability and irregularity, typically seen in disorganized or heterogeneous tissues, such as those associated with higher-grade tumors.

- Sum of Squares from the GLCM category (GLCM_SQ) Variance is a measure in the distribution of neigbouring intensity level pairs about the mean intensity level in the GLCM. It reflects the extent to which pixel intensities deviate from their average values across a given region of interest, meaning that higher GLCM_SQ values denote more variability and greater irregularity in tissue organization or architecture.
- Complexity from the NGTDM category (NGTDM_Com). The Complexity feature captures the degree of variation in the image. A higher Complexity value indicates more heterogeneity, while a lower value indicates more uniformity.
- Joint Energy from the GLCM category (GLCM_JE) quantifies the presence of homogeneous patterns in the image. Higher values imply more frequent occurrences of certain gray level pairs, indicating a more uniform and homogeneous texture.
- Inverse Difference Moment from the GLCM category (GLCM_IDM) measures the similarity of pixel intensities within a region; a high IDM value indicates low variability, high local consistency in pixel intensities, and greater homogeneity and uniformity.
- Correlation from the GLCM category (GLCM_Corr) evaluates the linear relationship between the gray levels of adjacent pixels. High correlation values signify strong linear dependence between pixel intensities, typically associated with more uniform textures and reduced heterogeneity.
- Entropy from the FO category (FO_E) specifies the uncertainty/randomness in the image values. Higher entropy values indicate greater randomness and higher heterogeneity in the pixel intensity distribution.
- Interquartile Range from the FO category (FO_IQR) represents the range between the 25th percentile (Q1) and the 75th percentile (Q3) of pixel intensity values within a tissue image. FO_IQR is valuable for assessing subtle changes in tissue variability, particularly in cases of moderate architectural disruption.
- Mean Absolute Deviation from the FO category (FO_MAD) is the mean distance of all intensity values from the Mean Value of the image array. Less deviation from the mean leads to more homeogenocity.
- Robust Mean Absolute Deviation from the FO category (FO_rMAD) is the mean distance of all intensity values from the Mean Value calculated on the subset of the image array with gray levels in between, or equal to the 10th and 90th percentile. FO_rMAD represents the structural variability or heterogeneity of tissue regions while avoiding distortion from extreme pixel values (outliers). Higher FO_rMAD values indicate greater variability and disordered or heterogeneous tissue structures. Lower FO_rMAD values suggest uniformity and consistent tissue patterns, typically associated with healthy or more organized tissue regions.
- Standard Deviation from the FO category (FO_SD) measures the amount of variation or dispersion from the Mean Value. By definition, standard deviation
- Skewness from the FO category (FO_Sk) measures the asymmetry of the distribution of values about the Mean value. Being symmetric to mean can show more homeogenocity.
- Kurtosis from the FO category (FO_Ku) measure of the 'peakedness' of the distribution of values in the image ROIs. Higher kurtosis means more pixels are similar to mean and more homeogenous.
- Variance from the FO category (FO_V) is the mean of the squared distances of each intensity value from the Mean value. This is a measure of the spread of the distribution of the mean. By definition, variance=$\sigma^2$. Higher variance means a higher difference from the mean and more heterogenocity.
- Uniformity from FO category (FO_Un) is a measure of the sum of the squares of each intensity value. This is a measure of the homogeneity of the image array. A greater uniformity implies a greater homogeneity or a smaller range of discrete intensity values. Based on the definition, greater uniformity implies greater homogeneity.
- Autocorrelation from the GLCM category (GLCM_AC) is a measure of the magnitude of the fineness and coarseness of texture. Higher autocorrelation indicates smooth and repetitive textures, suggesting uniformity in tissue structure. Lower autocorrelation reflects irregular or abrupt changes in intensity, often observed in cases of malignant or abnormal tumors.
- Cluster Prominence from the GLCM category (GLCM_CP) is a measure of the skewness and asymmetry of the GLCM. Higher values impliy more asymmetry about the mean, while a lower value indicates a peak near the mean value and less variation about the mean. Higher cluster prominence often

- suggests strong clustering of pixel intensities, indicating consistent or homogeneous tissue patterns often associated with normal or less-disrupted biological structures.
- Cluster Shade from the GLCM category (GLCM_CLS) is a measure of the skewness and uniformity of the GLCM. A higher cluster shade implies greater asymmetry about the mean. Higher GLCM_CLS values are associated with irregular textures suggesting disrupted tissue structure.
- Difference Entropy from the GLCM category (GLCM_DE) is a measure of the randomness/variability in neighbourhood intensity value differences. a higher Difference Entropy can indicate more complex, disordered tissue structures, such as fibrosis, inflammation, or malignancies.
- Joint Energy from the GLCM category (GLCM_JE) quantifies the presence of homogeneous patterns in the image. Higher values imply more frequent occurrences of certain gray level pairs, indicating a more uniform and homogeneous texture.
- Informational Measure of Correlation (IMC 1) from the GLCM category (GLCM_IMC1) assesses the correlation between the probability distributions of i and j, quantifying the complexity of the texture. High IMC1 values suggest strongly correlated textures, often indicative of organized tissue structures. For example, well-differentiated tissues may exhibit higher IMC1 due to uniform arrangements of cells and extracellular matrix.
- Informational Measure of Correlation (IMC 2) from the GLCM category (GLCM_IMC2) assesses the correlation between the probability distributions of I and j, quantifying the complexity of the texture. The range of IMC2 = [0, 1), with 0 representing the case of 2 independent distributions (no mutual information) and the maximum value representing the case of 2 fully dependent and uniform distributions (maximal mutual information, equal to $\log_2(N_g)$). High IMC 2 values (close to 1) suggest strong correlation and dependency between gray levels, indicating a more uniform and predictable texture. This corresponds to low heterogeneity, as the pixel values and their spatial relationships are consistent and less varied.
- Inverse Difference Moment from the GLCM category (GLCM_IDM) is a descriptor of tissue texture smoothness and homogeneity. IDM (a.k.a Homogeneity 2) is a measure of the local homogeneity of an image. IDM weights are the inverse of the Contrast weights (decreasing exponentially from the diagonal i=j in the GLCM). This feature assesses the similarity of pixel intensities within a region, a high IDM value means the region has low variability and high local similarity in pixel intensities and high homogeneity and uniformity.
- Maximal Correlation Coefficient (MCC) is a measure of complexity of the texture and $0 \leq MCC \leq 1$.
  In the case of a flat region, each GLCM matrix has a shape (1, 1), resulting in just 1 eigenvalue. In this case, an arbitrary value of 1 is returned. High MCC Values Indicate a strong correlation between gray levels.
  Reflect low heterogeneity, as the pixel intensities are more uniform and predictable.
- Size-Zone Non-Uniformity from the GLSZM category (GLSZM_SZN) measures the variability of Size Zone Volumes in the image, which a lower value indicating more homogeneity in size zone volumes. A lower SZN value indicates that the size zones are more uniform in volume. Relationship to Homogeneity: More uniform size zones suggest higher homogeneity. This is because a consistent size zone volume across the image means that the texture does not vary in terms of the size of the uniform regions, leading to a more homogeneous appearance.
- Size-Zone Non-Uniformity from the GLSZM category (GLSZM_SZN) quantifies the heterogeneity of zone sizes within a segmented region (like a tumor). A higher GLSZM_SZN value indicates greater variation in the sizes of zones with the same gray level, suggesting a more heterogeneous texture. Conversely, a lower value implies more uniformity in zone sizes. Low SZN values indicate that size zones in the tissue are consistently distributed and uniform, which may correspond to organized and homogeneous tissue structures. High SZN values signify significant variability in zone sizes, often indicating diseased states where structures are irregular, such as tumors or fibrosis.
- Large Area Emphasis from the GLSZM category (GLSZM_LAE) measures the prominence of large zones within the image texture. A higher GLSZM_LAE value indicates that there are larger zones present in the image, suggesting a texture with a tendency towards larger, more homogeneous regions. Conversely, a lower value suggests smaller, more fragmented zones, implying a more fine-grained texture.
- Size-Zone Non-Uniformity Normalized from the GLSZM category (GLSZM_SZNN) measures the variability of size zone volumes throughout the image, with a lower value indicating more homogeneity among zone size volumes in the image. This is the normalized version of the SZN formula.

- Gray Level Variance from the GLSZM category (GLSZM_GLV) measures the variance in gray level intensities for the zones.
- Zone Entropy (ZE) from the Gray-Level Size Zone Matrix (GLSZM) quantifies the uncertainty or randomness in the distribution of zone sizes and gray levels within an image. It measures the complexity or irregularity of texture patterns. Higher ZE values indicate greater heterogeneity and complexity in the texture patterns.
- Gray Level Non-Uniformity Normalized from the GLSZM category (GLSZM_GLNN) measures the variability of gray level intensity values in the image, with a lower value indicating a greater similarity in intensity values. This is the normalized version of the GLN formula.
- Gray Level Non-Uniformity from the GLSZM category (GLSZM_GLN) measures the variability of gray level intensity values in the image, with a lower value indicating more homogeneity in intensity values.
- Contrast from the NGTDM category (NTGDM_Co) is a measure of the spatial intensity change, but is also dependent on the overall gray level dynamic range. Contrast is high when both the dynamic range and the spatial change rate are high, i.e. an image with a large range of gray levels, with large changes between voxels and their neighborhood. Higher contrast values indicate a greater level of heterogeneity in the image, meaning there are significant differences in gray levels between adjacent regions. On the other hand, lower contrast values would suggest more uniformity in the distribution of gray levels across the image.
- Busyness from the NGTDM category (NTGDM_B) is a measure of the change from a pixel to its neighbor. A high value for busyness indicates a 'busy' image, with rapid changes of intensity between pixels and its neighborhoods. higher values indicating more rapid changes and lower values indicating more uniform, homogeneous regions. This feature is useful in characterizing the texture and identifying regions with high spatial frequency variations in medical images, such as areas with a lot of structural detail or noise.
- Large Dependence Emphasis (LDE) is a measure of the distribution of large dependencies, with a greater value indicative of larger dependence and more homogeneous textures. LDE provides a measure of the distribution of large, homogeneous dependencies in the image, which can be useful for texture analysis and characterization.
- Small Dependence Emphasis (SDE) is a measure of the distribution of small dependencies, with a greater value indicative of smaller dependence and less homogeneous texture. A higher SDE value indicates the presence of smaller, less homogeneous regions in the image texture.
- Small Dependence Low Gray Level Emphasis (SDLGLE) Measures the joint distribution of small dependence with lower gray level values. low intensity regions are within the image, reflecting the complexity and heterogeneity of the texture. This feature can be useful for distinguishing different tissue types or characterizing tumor heterogeneity in medical images
- Small Dependence High Gray Level Emphasis from the GLDM category (GLDM_SDHGLE) Measures the joint distribution of small dependence with higher gray level values. SDHGLE provides a measure of how prevalent small, high intensity regions are within the image, reflecting the complexity and heterogeneity of the texture. This feature can be useful for distinguishing different tissue types or characterizing tumor heterogeneity in medical images.
- High Gray Level Run Emphasis from the GLRLM category (GLRLM_HGLRE) measures the distribution of the higher gray level values, with a higher value indicating a greater concentration of high gray level values in the image. Higher HGLRE values indicate the dominance of bright, high-intensity areas forming longer runs within an image, signifying a well-defined structural feature.
- Gray Level Variance from the GLSZM category (GLSZM_GLV) measures the variance in gray level intensities for the zones. High GLSZM_GLV values reflect a greater range or spread of gray-level intensities within size zones, indicating significant tissue heterogeneity. Low GLSZM_GLV values indicate more uniform gray levels, which corresponds to homogeneous tissue structures.
- Coarseness from the NGTDM category (NGTDM_Coar) is a measure of average difference between the center voxel and its neighbourhood and is an indication of the spatial rate of change. A higher value indicates a lower spatial change rate and a locally more uniform texture. Coarseness measures the average difference between the center voxel and its neighborhood, indicating the spatial rate of change in the texture. A higher coarseness value suggests a lower rate of spatial change and a more uniform texture within ROIs. This means that areas with high coarseness have fewer abrupt intensity changes, reflecting a smoother and more homogeneous appearance.

- Gray Level Non-Uniformity from the GLDM category (GLDM_GLN) Measures the similarity of gray level intensity values in the image, where a lower GLN value correlates with a greater similarity in intensity values.
- Cluster Tendency from the GLCM category (GLCM_CT) is a measure of groupings of voxels with similar gray level values. A higher Cluster Tendency value indicates a stronger tendency for clusters of voxels with similar gray-level values to form within the image. This can imply that there are regions within the image where the gray-level values are more uniform or consistent, leading to the formation of clustered areas with similar texture patterns. Therefore, while Cluster Tendency does not directly measure uniformity, it is related to the concept of uniformity in terms of how it reflects the tendency of voxels to cluster together based on their gray-level values, suggesting regions of the image where the values are more consistent or uniform.

### 3.2. Semantic feature interpretability by PFs
### 3.2.1 Architectural Features

Although not directly mentioned in the WHO grading System, architectural features play a crucial role in understanding lesion characteristics. Textural features are more important in examining architecture. Signal intensity patterns are typically categorized into three distinct types: homogeneity, heterogeneity, and uniformity. Homogeneity, heterogeneity, and uniformity describe signal intensity patterns. Homogeneity indicates consistency, heterogeneity reflects variability (fine or coarse), and uniformity emphasizes even signal distribution, all critical for interpreting architectural features. Homogeneity refers to a consistent and uniform signal intensity throughout a region, indicating that the tissue or lesion is made up of similar components or structures. Features like GLDM_LDHGLE and GLDM_LDE highlight homogeneity in high values and coarse heterogeneity in low values, focusing on larger structures. Conversely, features like GLSZM_SAHGLE and GLRLM_SRLGLE emphasize fine heterogeneity, capturing smaller, non-uniform structures. Additionally, general features like FO_SD represent overall heterogeneity by measuring the variation or dispersion in intensity values across the texture. By contrast, heterogeneity refers to a structure with dissimilar components or elements, appearing irregular or variegated. Heterogeneity denotes variability in the signal, where intensity values fluctuate across the image. This can either be fine or coarse, depending on the level of variation. Coarse heterogeneity may suggest larger, more significant structural differences, while fine heterogeneity points to smaller, subtler variations within the tissue. As shown in Supplemental File; Table S5, homogeneity and heterogeneity visual semantic features, along with their associated PFs, are utilized to evaluate architectural patterns, with detailed relationships elaborated in Supplemental Sections 3,4,5. They can used for assessment of intranuclear homogeneity, heterogeneity, and uniformity (fine or coarse).

### 3.2.2 Cytologic Features

Cell detection modules are specifically related to cell size (including nuclear size, cytoplasmic size, and nuclear to cytoplasmic ratio) and shape (circularity and caliper). Terminologies that favor malignancy in pathology are mostly based on nuclear features such as pleomorphism, atypia, nuclear size, and irregularity. In this regard, features such as MEAN Nucleus Area (MNA), MEAN Nucleus Perimeter (MNP), MEAN Nucleus Circularity (MNC), MEAN Nucleus Max Caliper (MNMxC), MEAN Nucleus Min Caliper (MNMnC), and MEAN Nucleus Eccentricity (MNE), can be indicative of nuclear features. A A fewfeatures are indicative of cellular atypia, which are described in detail in the Supplemental File; Table S2. Nucleus Area in definition means the average size (area) of cell nuclei, expressed in square micrometers (µm²) or pixels. The higher the area of the nucleus indicates the atypia of the nucleus. Of course, it is better to interpret this feature together with the nucleus to the cytoplasm ratio (Nucleus Cell Area Ratio). The perimeter of the nucleus (name as Nucleus Perimeter) is a measure of the nucleus size in HCC. Nuclear irregularity, including changes to the perimeter, reflects atypical growth and malignancy. Larger or more irregular nuclear perimeters are often associated with higher-grade tumors or more aggressive subtypes, as they may indicate abnormal cell division and altered growth patterns. Nucleus Eccentricity measures the degree to which a shape deviates from being circular, with higher values representing more elongated shapes. It is defined as the ratio of the distance between the foci of an ellipse to its major axis length. In a perfect circle, the eccentricity is 0. The value ranges from 0 to 1: an eccentricity of 0 indicates a perfect circle, while values closer to 1 indicate increasingly elongated shapes, which are typically observed in tumor cells undergoing atypical or more aggressive transformations

    Features that measure color intensity described as MEAN Nucleus Hematoxylin Optical Density (MNHODm), MEAN Nucleus Hematoxylin Optical Density Sum (MNHODS), MEAN Nucleus Hematoxylin Optical Density Standard Deviation (MNHODD), MEAN Nucleus Hematoxylin Optical Density Max (MNHODMx), MEAN Nucleus Hematoxylin Optical Density Min (MNHODMn), MEAN Nucleus Hematoxylin Optical Density Range (MNHODR), MEAN

Nucleus Eosin Optical Density Mean (MNEODm), MEAN Nucleus Eosin Optical Density Sum (MNEODs), MEAN Nucleus Eosin Optical Density Standard Deviation (MNEODSD), MEAN Nucleus Eosin Optical Density Max (MNEODMx), MEAN Nucleus Eosin Optical Density Min (MNEODMn), and MEAN Nucleus Eosin Optical Density Range (MNEODR) and extended relationship with WHO grading in Supplemental Table S5. Additionally, following PFs highlight both the homogeneity and intensity of tumoral cells, as these substantial regions significantly influence the lesion's overall characteristics. Well-differentiated lesion usually had uniform cells with moderate color intensity. In contrast, lesions with poorly differentiated features are associated with WHO Grade more than 2. All PFs such as Large Dependence Low Gray Level Emphasis(LDLGLE) and Large Dependence High Gray Level Emphasis (LDHGLE) from the GLDM category (GLDM_LDLGLE and GLDM_LDHGLE), Large Area High Gray Level Emphasis and Large Area Low Gray Level Emphasis from the GLSZM category (GLSZM_LAHGLE and GLSZM_LALGLE), Long Run High Gray Level Emphasis from the GLRLM category (GLRLM_LRHGLE), and the Long Run Low Gray Level Emphasis from the GLRLM category (GLRLM_LRLGLE) are associated with intensity characteristics, as outlined in Supplemental File,Table S1 and elaborated in Supplemental File; Section 5.

### 3.3. Interpretable and Explainable Classification Task for Understandable AI

This study utilized open-source QuPath software extensions to apply a range of features to publicly available liver cancer datasets from the National Cancer Institute. The primary goal was to create a Practical Pathobiological Dictionary that links visual semantic features commonly used in clinical diagnostics with established risk factors and predictive markers for liver cancer. This dictionary provides a standardized framework to interpret PFs derived from pathology slides, enabling clinicians to bridge the gap between complex AI outputs and practical clinical workflows. To ensure clinical relevance, a combination of both simple and complex classification models was employed, with FSAs used to identify critical features related to liver cancer prognosis. These tools helped validate predictive models by enhancing interpretability, ensuring transparency, and maintaining relevance to clinical practice. The cohort consisted of 60.4% men and 39.6% women, with an average patient age of approximately 33.89 years. As shown in Figure 4, combining quantitative features with VTS + SVM achieved the highest average accuracy of 0.80±0.01, with an external nested testing accuracy of 0.80±0.02 (Paired-t test,P-value <0.05). As shown in Table 1 the top 20 selected features from nine FSA Models (Totally 90 features) include 47 PFs (such as Centroid Y μm, Nucleus: Hematoxylin Optical Density min (HODMn), other depicted in Table1), 32 variables from outcome measure table in the dataset (such as Age, Prior treatment before surgery, AJCC pathologic stage, Prior malignancy before HCC, Year of diagnosis, Race, Tumor grade and Treatment type), and 11 texture features (such as GLDMl_Dependence Non Uniformity (GLDM_DN), GLDM_Gray Level Non Uniformity (GLDM_GLN), GLRLM_Gray Level Non Uniformity (GLRLM_GLN), GLRLM_ Long Run Emphasis (GLRLM_LRE), GLRLM_Long Run High Gray Level Emphasis (GLRLM_LRAHGLE), GLRLM_Long Run Low Gray Level Emphasis (GLRLM_LRLGLE) _Run Length Non Uniformity (GLRLM_RLN), GLRLM_Run Variance (GLRLM_RV), GLSZM_Large Area Emphasis (GLSZM_LAE), GLSZM_Large Area High Gray Level Emphasis (GLSZM_LAHGLE), GLSZM_Large Area Low GrayLevel Emphasis (GLSZM_LALGLE)).

Moreover, some ML algorithms reached a performance of around 0.73. These FSAs selected mainly consist of nuclear and cytoplasmic features referring to nuclear atypia, pleomorphism, and hyperchromasia. Here, we explain the top 20 features with the highest accuracy related to overall survival, specifically in the VTS+ SVM algorithm, and examine their relationship with semantic features. We provide detailed explanations about the main features of other algorithms in the supplemental section 3. Utilizing a range of smoothed measures—specifically 50, 100, 150, and 200 μm—proved beneficial in our selection model for two main reasons: to analyze the selected features and to assess details effectively. In image processing, "Smoothed: 50" pertains to a filter with a smaller kernel size, enabling retention of more detail and sharpness, as it smooths over a limited area of pixels. Conversely, "Smoothed: 200" employs a larger smoothing value, which significantly reduces detail by averaging over a broader area, resulting in a more blurred image. This creates a notable trade-off between detail retention and noise reduction; smoothing at a value of 50 preserves more details (such as edges and textures) while smoothing at 200 can eliminate unwanted noise, albeit at the expense of losing finer details and structural integrity. Of the 20 features selected in VTS, 12 were clinical and demographic features, while the rest were related to cell detection features (PFs):

- Feature 1; Cytoplasm Eosin Optical Density max (CYEODMx measures the OD of eosin staining, which reflects the cytoplasmic protein content and tumor cell differentiation. A higher eosin OD may indicate cells with more dense protein and specific morphology, often correlating with the differentiation level. In confirmation of this issue, by referring to the WHO grading System (see Supplemental Table S3), we find that in cases of well-to-moderate differentiation, the cytoplasm is defined as Eosinophilic to basophilic cytoplasm. As cellular differentiation decreases and the grade increases, the cytoplasm becomes more basophilic. Thus, CYEODMx can serve as a quantitative proxy for assessing cellular differentiation in HCC tissues, aiding in accurate tumor grading.

- Feature 2; CYEODMx (Smoothed: 50 µm) represents the maximum OD of eosin staining within cytoplasm regions, but it has undergone a smoothing process with a window size of 50 µm. Smoothing is typically applied to reduce noise or variability in the data, creating a more generalized measurement over spatial scales. CYEODMx (Smoothed, 50 µm) utilizes a smoothing technique over a 50 µm scale to give a less detailed but potentially more consistent value, whereas the CYEODMx is more localized and may capture finer variations.
- Feature 3; Nucleus: Eosin Optical Density mean (NEODm) (Smoothed: 50 µm) represents the average optical density of eosin staining in nuclear regions, calculated after applying a spatial smoothing window of 50 µm. It provides a general measure of how strongly eosin stains the nuclei, with reduced impact from fine-grained variability.
- Feature 4; NEODmrepresents the average level of eosin staining (quantified by OD) within the nuclear areas of cells. While the nucleus is typically stained more prominently by hematoxylin, this feature measures the subtle eosinophilic staining present in the nucleus. It might provide insights into nuclear structure or eosinophilic content in specific biological contexts.
- Feature 5; Prior Treatment Before Surgery: This feature refers to any therapies (e.g., local treatments like radiofrequency ablation, transarterial chemoembolization (TACE), systemic treatments, or targeted therapy) that the patient received before surgery to manage the tumor. Although prior treatment does not directly affect the WHO grade (which focuses on histology), it can have indirect effects. Treatments might alter the tumor's appearance, making grading more challenging or influencing differentiation. It may affect tumor vascularization, necrosis, or size, which might correlate with aggressiveness. Prior treatments may reflect tumor complexity (e.g., larger or multifocal tumors requiring intervention), generally linked to more advanced disease.
- Feature 6; NEODm [Smoothed: 100µ] represents the average level of eosin staining (quantified as OD) within nuclear regions after applying spatial smoothing with a window of 100 µm. This feature can be used to assess subtle eosinophilic characteristics of nuclei, possibly related to specific biological or pathological conditions, while removing fine-grained variations in staining intensity.
- Feature 7; NEODm [Smoothed: 200 µm] represents the average intensity of eosin staining within the nuclei, calculated after applying spatial smoothing over a scale of 200 µm. This feature helps to analyze the overall distribution and concentration of eosin staining in nuclear regions at a broader scale, offering insights into tissue morphology and potential pathological characteristics while minimizing localized noise and variability.
- Feature 8; NEODm [Smoothed: 150 µm] represents the average eosin staining intensity (optical density) within nuclear regions, computed after applying a smoothing process over a 150 µm spatial scale. This feature provides a generalized measure of eosinophilic staining in nuclei, which may offer insights into tissue structure or pathological changes while minimizing localized variability and noise.
- Feature 9; Age is often correlated with disease outcome. For example, younger patients may show rapid tumor progression, while older patients may have slower disease progression depending on associated risk factors (e.g., comorbidities). HCC typically occurs in middle-aged or older individuals due to risk factors like chronic hepatitis B/C or alcohol-related cirrhosis, which take years to develop.
- Feature 10; AJCC Pathologic stage classification_Tumor: The AJCC Pathologic Tumor Classification evaluates the size, local invasion, and PFs of the primary tumor as part of the TNM system. For HCC, it specifically examines whether the tumor is solitary, multiple, has vascular invasion, or invades nearby structures (T1–T4).
- Feature 11; Prior malignancy: In the TCGA-LIHC dataset, the prior_malignancy row refers to whether a patient has a history of malignancy (cancer) before the current diagnosis of HCC. A history of prior malignancy may impact survival rates, treatment options, and recurrence risk.
- Feature 12; Year of diagnosis: In the TCGA-LIHC - LIHC) dataset, the year_of_diagnosis row refers to the calendar year when the patient was clinically diagnosed with HCC. Being diagnosed earlier may indicate fewer advanced treatment options, while diagnosis in later years may reflect newer therapeutic approaches.

- Feature 13; Centroid X µm: ML algorithms in pathology utilize centroid measurements to enhance diagnostic accuracy, automating the identification of abnormal tissues or cells. Examining the centroids of tumor cells is crucial for assessing heterogeneity within the tumor microenvironment, which can influence prognosis and treatment decisions. Centroid X µm represents the horizontal (X-axis) position of the geometric center (centroid) of an object, such as a cell or nucleus, with the distance measured in micrometers. It is used in histological and spatial analyses to map and study the arrangement of structures within a tissue sample.
- Feature 14: Race: Analyzing this variable can help researchers examine potential health disparities in HCC incidence, progression, treatment response, and survival outcomes across different racial groups. Genetic, socioeconomic, and environmental factors may contribute to these disparities. While not directly related to race, there is population-level genetic variation that needs to be considered in research. Researchers must account for confounding factors to avoid misinterpreting correlations between race and other variables. It's crucial to remember that race is a social construct, and observed differences may be related to other factors like access to healthcare, socioeconomic status, lifestyle, and environmental exposures, rather than inherent genetic differences.
- Feature 15; Tumor grade: indicates the level of differentiation of liver cancer cells compared to normal liver tissue. It provides crucial prognostic information about the aggressiveness of HCC, influencing the prognosis. The fact that tumor grade has been among the top 20 features in feature selection itself indicates that the process of selecting WHO grading for examining survival and the correlation between numerical features and semantic features has been relevant and logical.
- Feature 16; Treatment type: captures the category of treatment administered to the patient. The treatment type may pertain to local therapies, systemic therapies, surgical interventions, or supportive treatments. It is essential for understanding the relationship between treatment approaches, survival, tumor progression, and overall prognosis.
- Feature 17; AJCC Pathologic stage classification_Node: The AJCC Pathologic Stage Classification for Node (N) provides important information regarding the extent of regional lymph node involvement in patients with cancer, including HCC. The classification ranges from N0 (no lymph node involvement) to N3 (involvement of distant lymph nodes), impacting prognosis and treatment strategies significantly. Understanding the N classification is crucial for accurate staging, treatment planning, and predicting outcomes in cancer patients.
- Feature 18; AJCC Pathologic stage classification _Metastasis: The AJCC Pathologic Stage Classification for Metastasis (M) provides essential information regarding the presence and extent of distant metastasis in cancer patients, including those with HCC. The classification distinguishes between M0 (no distant metastasis) and M1 (distant metastasis present), which significantly impacts prognosis, treatment decisions, and overall disease management. Understanding the M classification is critical for precise staging, effective treatment planning, and predicting patient outcomes.
- Feature 19; Treatment or therapy after surgery captures whether a patient received additional therapy following surgery for HCC. This data is critical for understanding the role of post-surgical interventions in improving patient outcomes and preventing recurrence, as well as providing insights into treatment patterns in HCC management.
- Feature 20; Tumor grade influences HCC through epidemiological, biological, hormonal, and behavioral mechanisms. Male patients are at higher risk due to androgen-driven carcinogenesis, lifestyle factors, and increased susceptibility to liver injury. Female patients benefit from the protective effects of estrogens, stronger immune responses, and slower disease progression, often resulting in better prognosis. Gender differences highlight the need for personalized approaches in the prevention, treatment, and management of HCC.

**Table. 1.** Selected features by different Feature Selection Algorithms (FSAs) applied on Qupath/PyRadiomics extracted features. Each column represents the selected features by FSAs.

| Feature Selector | Chi-Square Test (CST) | Correlation Coefficient (CC) | Mutual Information (MIS) | Variance Threshold (VTS) | ANOVA F-test (AFT) | Information Gain (IGS) | Univariate Feature Selection (UFS) | Fisher Score (FSF) | Least Absolute Shrinkage and Selection Operator (LASSO) |
|---|---|---|---|---|---|---|---|---|---|
| Feature 1 | Centroid Y µm | Age | Nucleus: Hematoxylin Optical Density min | Cytoplasm: Eosin Optical Density max | Centroid Y µm | Nucleus: Hematoxylin Optical Density min | Nucleus: Hematoxylin OD min | Age | Centroid Y µm |
| Feature 2 | Nucleus: Circularity | Smoothed: 100 µm: Nucleus: Circularity | Cell: Perimeter | Smoothed: 50 µm: Cytoplasm: Eosin Optical Density max | Nucleus: Circularity | Cell: Perimeter | Cell: Perimeter | Smoothed: 100 µm: Nucleus: Circularity | Nucleus: Eosin Optical Density range |
| Feature 3 | Smoothed: 50 µm: Nucleus: Circularity | Smoothed: 200µm: Nucleus: Hematoxylin Optical Density range | Smoothed: 50 µm: Cell: Perimeter | Smoothed: 50 µm: Nucleus: Eosin Optical Density mean | Cell: Hematoxylin Optical Density min | Smoothed: 50µm: Cell: Perimeter | Smoothed: 50 µm: Cell: Perimeter | Smoothed: 200 µm: Nucleus: Hematoxylin Optical Density range | Smoothed: 100 µm: Nucleus: Max caliper |
| Feature 4 | Smoothed: 100 µm: Nucleus: Circularity | Smoothed: 200 µm: Nucleus: Circularity | Smoothed: 50 µm: Nucleus/Cell area ratio | Nucleus: Eosin Optical Density mean | Cytoplasm: Hematoxylin Optical Density min | Smoothed: 100 µm: Nucleus: Circularity | Smoothed: 100 µm: Nucleus: Circularity | Smoothed: 200 µm: Nucleus: Circularity | Smoothed: 100 µm: Nucleus: Eccentricity |
| Feature 5 | Smoothed: 150 µm: Nucleus: Circularity | Smoothed: 150 µm: Nucleus: Circularity | Smoothed: 100 µm: Nucleus: Circularity | Prior treatment before surgery | Smoothed: 50 µm: Cell: Hematoxylin Optical Density min | Smoothed: 100 µm: Nucleus: Hematoxylin Optical Density std dev | Smoothed: 100 µm: Nucleus: Hematoxylin Optical Density std dev | Smoothed: 150 µm: Nucleus: Circularity | Smoothed: 100µm: Nucleus: Hematoxylin Optical Density mean |
| Feature 6 | Smoothed: 200 µm: Nucleus: Circularity | AJCC_metastasis | Smoothed: 100 µm: Nucleus: Hematoxylin OD std dev | Smoothed: 100µm: Nucleus: Eosin Optical Density mean | Smoothed: 50 µm: Cytoplasm: Hematoxylin Optical Density min | Smoothed: 100 µm: Nucleus: Hematoxylin Optical Density min | Smoothed: 100 µm: Nucleus: Hematoxylin Optical Density min | AJCC_metastasis | Smoothed: 100 µm: Nucleus: Hematoxylin OD sum |
| Feature 7 | Nucleus: Solidity | Nucleus: Circularity | Smoothed: 100 µm: Nucleus: Hematoxylin OD min | Smoothed: 200 µm: Nucleus: Eosin Optical Density mean | Smoothed: 100 µm: Nucleus: Circularity | Smoothed: 100 µm: Cell: Area | Smoothed: 100 µm: Cell: Area | Nucleus: Circularity | Smoothed: 100 µm: Nucleus: Hematoxylin OD Std Dev |
| Feature 8 | original_GLDM_DependenceNonUniformity (GLDM_DN) | Centroid Yµm | Smoothed: 100 µm: Cell: Area | Smoothed: 150 µm: Nucleus: Eosin Optical Density mean | Smoothed: 100 µm: Cell: Hematoxylin Optical Density min | Smoothed: 100 µm: Cell: Perimeter | Smoothed: 100 µm: Cell: Perimeter | Centroid Y µm | Smoothed: 100 µm: Nucleus: Hematoxylin OD Max |
| Feature 9 | original_ GLDM _GrayLevelNonUniformity((GLDM_GLN) | Cell: Hematoxylin Optical Density min | Smoothed: 100 µm: Cell: Perimeter | Age | Smoothed: 100 µm: Cytoplasm: Hematoxylin Optical Density min | Smoothed: 100 µm: Cell: Max caliper | Smoothed: 100 µm: Cell: Max caliper | Cell: Hematoxylin OD Min | Smoothed: 100 µm: Nucleus: Hematoxylin Optical Density Min |
| Feature 10 | original_GLRLM_GrayLevelNonUniformity(GLRLM_GLN) | Cytoplasm: Hematoxylin Optical Density min | Smoothed: 100 µm: Cell: Max caliper | AJCC_tumor | Smoothed: 150 µm: Nucleus: Circularity | Smoothed: 150 µm: Nucleus: Circularity | Smoothed: 150 µm: Nucleus: Circularity | Cytoplasm: Hematoxylin Optical Density Min | Smoothed: 100 µm: Nucleus: Hematoxylin Optical Density Range |
| Feature 11 | original_GLRLM_LongRunEmphasis (GLRLM_LRE) | Smoothed: 50 µm: Cell: Hematoxylin Optical Density min | Smoothed: 150 µm: Nucleus: Circularity | Prior malignancy | Smoothed: 150 µm: Cell: Hematoxylin Optical Density min | Smoothed: 150 µm: Nucleus: Hematoxylin Optical Density std dev | Smoothed: 150 µm: Nucleus: Hematoxylin Optical Density std dev | Smoothed: 50 µm: Cell: Hematoxylin Optical Density Min | Smoothed: 100 µm: Nucleus: Eosin Optical Density Mean |
| Feature 12 | original_GLRLM_LongRunHighGrayLevelEmphasis (GLRLM_LRAHGLE) | Smoothed: 100 µm: Cell: Hematoxylin Optical Density min | Smoothed: 150 µm: Nucleus: Hematoxylin OD std dev | Year of diagnosis | Smoothed: 150 µm: Cytoplasm: Hematoxylin Optical Density min | Smoothed: 150 µm: Cell: Area | Smoothed: 150 µm: Cell: Area | Smoothed: 100 µm: Cell: Hematoxylin Optical Density Min | Smoothed: 100 µm: Nucleus: Eosin Optical Density Sum |
| Feature 13 | original_GLRLM_LongRunLowGrayLevelEmphasis GLRLM_LRLGLE) | Smoothed: 50 µm: Cytoplasm: Hematoxylin Optical Density min | Smoothed: 150 µm: Cell: Area | Centroid X µm | Smoothed: 200 µm: Nucleus: Circularity | Smoothed: 150 µm: Cell: Perimeter | Smoothed: 150 µm: Cell: Perimeter | Smoothed: 50 µm: Cytoplasm: Hematoxylin Optical Density min | Smoothed: 100 µm: Nucleus: Eosin Optical Density Std Dev |
| Feature 14 | original_GLRLM_RunLengthNonUniformity (GLRLM_RLN) | Smoothed: 150 µm: Cell: Hematoxylin Optical Density min | Smoothed: 150 µm: Cell: Perimeter | Race | Smoothed: 200 µm: Nucleus: Hematoxylin Optical Density range | Smoothed: 150 µm: Cell: Max caliper | Smoothed: 150 µm: Cell: Max caliper | Smoothed: 150 µm: Cell: Hematoxylin Optical Density Min | Smoothed: 100 µm: Nucleus: Eosin Optical Density max |
| **Feature 15** | original_GLRLM_RunVariance(GLRLM_RV) | Smoothed: 100 µm: Cytoplasm: Hematoxylin Optical Density min | Smoothed: 150 µm: Cell: Max caliper | Tumor grade | Smoothed: 200 µm: Cell: Hematoxylin Optical Density min | Smoothed: 200µm: Cell: Area | Smoothed: 200 µm: Cell: Area | Smoothed: 100 µm: Cytoplasm: Hematoxylin Optical Density Min | Smoothed: 100 µm: Nucleus: Eosin Optical Density min |
| **Feature 16** | original_GLSZM_LargeAreaEmphasis | Smoothed: 200 µm: Cell: Hematoxylin Optical Density min | Smoothed: 200 µm: Cell: Area | Treatment type | Smoothed: 200 µm: Cytoplasm: Hematoxylin Optical Density min | Cell: Length µm | Cell: Length µm | Smoothed: 200 µm: Cell: Hematoxylin Optical Density Min | Nucleus: Solidity |
| **Feature 17** | original_GLSZMLargeAreaHighGrayLevelEmphasis | Smoothed: 150 µm: Cytoplasm: Hematoxylin Optical Density min | Smoothed: 200 µm: Cell: Max caliper | AJCC_Node | Nucleus: Solidity | Delaunay: Mean distance | Delaunay: Mean distance | Smoothed: 150 µm: Cytoplasm: Hematoxylin Optical Density min | Age |
| **Feature 18** | original_GLSZM_LargeAreaLowGrayLevelEmphasis | Smoothed: 200 µm: Cytoplasm: Hematoxylin Optical Density min | Cell: Length µm | AJCC_metastasis | Age | original_GLRLM_RunVariance((GLRLM_RV) | original_GLRLM_RunVariance | Smoothed: 200 µm: Cytoplasm: Hematoxylin Optical Density min | Race |
| **Feature 19** | AJCC_metastasis | Nucleus: Solidity | Delaunay: Mean distance | Treatment or therapy after surgery | AJCC_metastasis | Age | Age | Nucleus: Solidity | AJCC _pathologic_m |
| **Feature 20** | Tumor grade | Tumor grade | Age | Gender | Tumor grade | Treatment type | Treatment type | Tumor grade | Tumor grade |

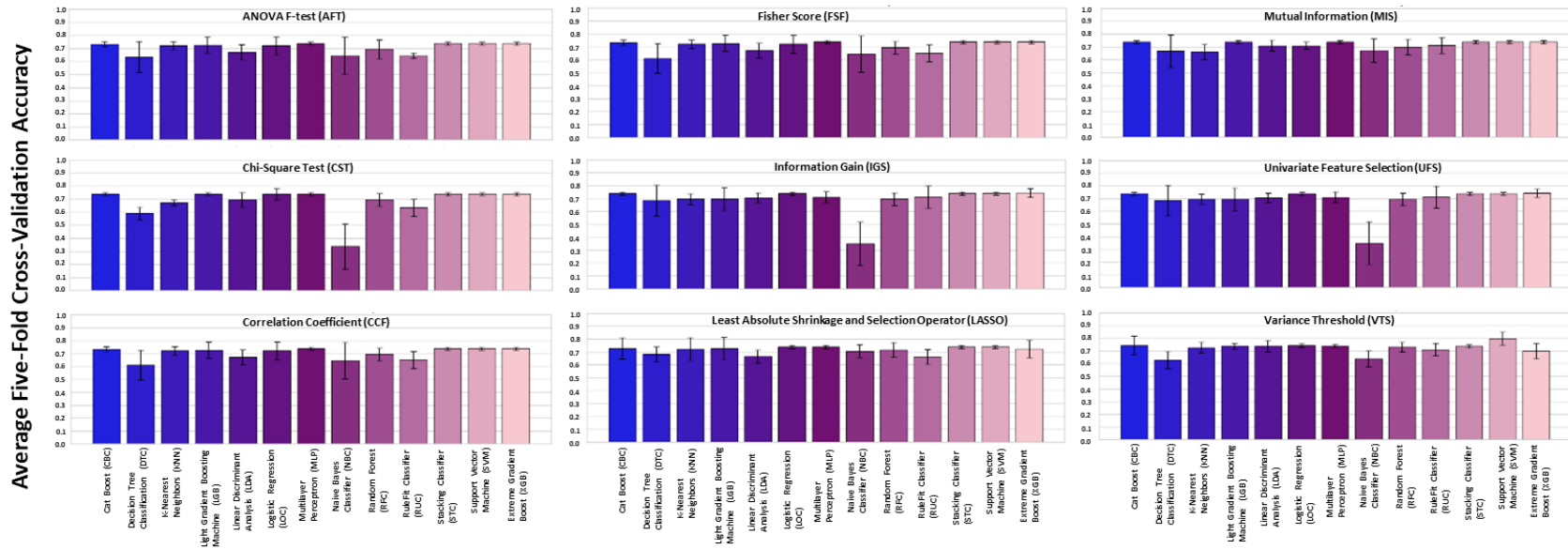

**Fig. 4.** Bar plot of Mean ± standard deviation of Feature Selection Algorithms (FSA) applied to classification algorithms (CA). This figure showcases the average five-fold cross-validation accuracy of 13 CAs applied to feature subsets derived from nine different FSAs. Each subplot represents the performance of CAs linked with feature selection algorithms. Bars indicate the mean validation accuracy, with error bars representing variation across repeated runs, providing insights into the predictive performance and stability of the classifiers for liver cancer classification across various feature subsets.

## 4. DISCUSSION

In this dictionary, we aimed to utilize a reliable metric, such as the WHO Grading System, to help researchers link PFs to semantic features and understand their clinical definitions. This approach was designed to enhance clinical relevance and improve the transparency of decision-making in AI models. Furthermore, by testing a series of features and examining their correlation with overall survival, we sought to validate our findings. Our analysis of FSAs clearly demonstrated that specific demographic and clinical attributes are highly influential factors in outcomes such as overall survival, underscoring the critical impact of these variables. We interpreted the results keeping in mind that the QuPath has limitations; Overlapping or clustered nuclei are often mis segmented, Parameters (e.g., nucleus size, threshold) need fine-tuning for each slide type, especially across different staining or tissue types.

The most critical features were related to pathomics characteristics derived from cell detection, emphasizing cellular traits such as the nucleus and cytoplasm. This is particularly important as, to date, these cellular attributes have not been sufficiently examined in relation to their definition and correlation with semantic features of grading for specific cancers. This insight could guide future work in this area. Cellular features, especially those related to the nucleus and cytoplasm, were identified as highly impactful in predicting patient survival. Since these features are tied to grading, they also indirectly underscore grading as a predictor of patient survival outcomes. Subsequently, we specifically examined the algorithm with the highest accuracy, VTS+SVM, which achieved the highest average accuracy of $0.80\pm0.01$. We analyzed its 20 selected features in detail and explained those that could be linked to semantic features or patient survival. Among these, the majority were clinical attributes whose importance in predicting patient prognosis has already been established in the literature. Other features, such as nuclear features including staining, size, and shape, also emerged as critical findings in our results. Interestingly, none of the texture features were present in the VTS algorithm, as their aggregation was observed more in other algorithms. This exploration underscores the relevance of cellular features and their connection to clinical outcomes, providing a foundation for subsequent research in the field.

Recent advancements in computer vision and digital microscopy have revolutionized the landscape of cancer diagnosis, particularly in the domain of histopathological image analysis. The integration of AI into clinical practice has shown tremendous potential for enhancing diagnostic accuracy, improving workflow efficiency, and supporting decision-making processes. However, despite these benefits, the adoption of AI in healthcare faces significant challenges, primarily due to its "black box" nature. This inherent lack of transparency in many AI systems raises concerns among clinicians and healthcare stakeholders, as decisions made by AI algorithms can often seem opaque or difficult to interpret [90, 91]. Addressing these challenges requires the development of frameworks that prioritize explainability and causality, particularly for high-stakes applications like cancer diagnostics.

In the context of HCC diagnostics, bridging digital tools with standardized clinical frameworks has emerged as a promising approach. For instance, integrating PFs with WHO grading criteria has demonstrated the potential to improve both interpretability and diagnostic precision. By combining objective, quantitative metrics derived from histopathological images with qualitative clinical standards, AI-driven systems can deliver outputs that are both scientifically grounded and aligned with clinical best practices [92]. This dual approach not only enhances diagnostic reliability but also fosters clinician trust in AI systems.

To further advance the integration of AI into clinical workflows, there is an urgent need to establish shared definitions and standardized protocols that govern the use of AI in medicine. Key questions must be addressed: What constitutes AI in the context of histopathology? What specific algorithms are being used, and what are their limitations and advantages? How can they be applied in ways that maximize impact while addressing ethical considerations and practical constraints? Answering these questions will allow researchers and clinicians to align their understanding of AI's capabilities and limitations, thereby facilitating deployment in real-world settings.

Another critical component of this discussion lies in identifying stakeholders and their diverse objectives. Clinicians, researchers, patients, and policymakers each approach AI applications in healthcare with distinct goals and requirements. For example, clinicians may prioritize the interpretability and usability of AI tools in their diagnostic workflow, whereas researchers may focus on improving algorithmic accuracy and exploring novel features like pathomics data. Patients and policymakers, on the other hand, may emphasize ethical concerns, data security, and equitable access to AI technologies. A successful AI framework must incorporate all these perspectives, balancing technical efficiency and scientific rigor with ethical and practical considerations.

The impact of tumor grading on various clinical outcomes has been investigated in some studies and from a variety of aspects [93, 94, 95, 96, 87]. These include the impact of grading on response to treatment [94], recurrence after resection and liver transplantation [96], disease progression [95], Survival [93], and metastasis [87]. Han et al. [93] showed that being in the worst histological group (poorly differentiated) was identified as an independent and negative predictor of disease-free survival (DFS) and overall survival (OS). The 5-year survival rate and recurrence-free survival rate of HCC patients were significantly greater in cases in which the main nodule showed well-differentiated HCC than in other cases [95]. There were differences in surgical outcome in relation to the histopathological grade of differentiation of the main nodule in patients with synchronous multicentric HCC. In addition, it has been shown that poorly differentiated tumors are inversely related to tumor nodule size and directly related to alpha-fetoprotein levels [95]. moderately to poorly differentiated HCC tissue and are more frequently associated with portal vein invasion and intrahepatic metastasis than the simple nodular type [87]. We also demonstrated this in Feature selection analysis. The presence of the tumor grade feature (from the dataset) among the 20 selected features indicates that Tumor Grade is associated with patient outcome and could predict the survival outcome of patient (Section 3.4).

One possible solution is the creation of a standardized dictionary or knowledge base that defines key semantic terms, algorithms, and workflows in the domain of AI-driven histopathology. Such a resource would act as a bridge between different stakeholders, ensuring consistent terminology and fostering interdisciplinary collaboration. By mapping the connections between semantic features, PFs, and clinical outcomes, researchers can make AI systems more transparent, allowing stakeholders not only to understand the "how" of AI-driven diagnostics but also the "why." This would contribute to human-centered AI functions that prioritize explainability, ensuring these tools meet the demands of clinical use while retaining the confidence of those who rely on them.

The important issue regarding the structure of dictionaries is that the dictionary should be understandable for both doctors and programmers [97]. Doctors do not trust the algorithm because the features are not directly related to semantics, and on the other hand, programmers, due to their lack of knowledge in recognizing the semantics of the features, rely solely on ML methods. The existence of a gap in linking the features to the semantics is necessary through knowledge of pathology and by a pathologist familiar with ML led to the design of this study. This study [98] found that 73% of explainable AI studies do not involve clinicians during development, increasing the risk of producing results that lack clinical relevance. 87% of explainable AI studies do not assess the quality of their explanations, reducing their clinical utility and compromising trust among common user post-hoc explainable AI methods, such as SHAP or LIME, often rely on oversimplified assumptions that can lead to inaccuracies in explanations.

Explainability is one of the most important points to establish trustworthiness in AI generally and in computational pathology specifically. Although there is a complete definition of texture features in RFs [46], these definitions are not clearly adjusted in pathology [99]. However, some of the features defined in relation to image features are suited to pathology images. In pathological image analysis studies, although texture feature extraction is generally simpler than shape features, their interpretation is more ambiguous and may be less accurate, so we have limitations in interpreting texture features and giving them meaning with semantic features. Studies that extract features from a pathology slide using both RFs and PFs have been rare, and these have been limited to pathology software and on organs other than the liver [100]. Although we extracted texture features through Python to have a complete collection and even defined all of them with their formulas, its application to all pathological images is limited. Mainly, GLCM-related features are used more due to better accessibility and simpler interpretation [101].

The application of AI to assist pathologists in diagnosing HCC, coupled with quantitative measurements of tissue morphology, is an undeniable necessity. AI not only streamlines the diagnostic process, enabling pathologists to reach conclusions more rapidly, but also enhances diagnostic accuracy and consistency. Using deep learning to more accurately classify tumors like HCC requires semantic knowledge, which can be derived from dictionaries. Different studies [102, 103, 104, 105] attempt to segment nuclei, trabecular, and tissue change features to more accurately improve the classification task of deep learning methods but rare studies focus mainly on the semantic explainability of quantitative measures. For radio-pathomics studies, a dictionary is also essential because understanding the interrelationship between these two entities requires understanding the corresponding semantics between these two groups. It has even been found that a number of CT-derived RFs such as Median, Autocorrelation, Contrast, Low Gray Level Emphasis (LGLE), Dependence Entropy (DE), Dependence Non-Uniformity (DN), Large Dependence Emphasis (LDE), Coarseness, Elongation, and Flatness were valuable for noninvasively assessing the micro-vascular invasion and histopathologic grade of HCC [106]. It seems that the strength of PFs is greater in higher WHO grades. That is, the less similar tumoral tissue is to normal tissue, the greater the resolution of these features, especially texture features, and its due to the presence of noise. Especially in lower grades, where the tumoral tissue is similar to normal tissue, and the presence of fatty change, Mallory–Denk body (MDB) bodies and normal sinusoids in the tissue background can cause interference. A novel algorithm [107] was introduced to determine the number of trabecular cell layers in histological images of HCC stained with (HE). The proposed method

*Salmanpour et al.*                                              *Dictionary Version LCP1.0*involved constructing a Delaunay diagram based on nuclei identified through template matching, followed by the removal of sinusoidal regions and fat droplets from the diagram. A thinning algorithm was then applied to refine the analysis. The results demonstrated that the calculated number of cell layers varied significantly with both tumor differentiation and Edmondson grades, indicating its potential utility in HCC diagnostic support systems.

As features like hepatocyte nuclei and trabecular features of tumors have been utilized as one of the main sources of features used by pathologist in HCC diagnosis [102], we thoroughly addressed the Tumor microenvironment tissue characteristics by specific handcrafted annotation, and we dedicate research on the peritumoral and intranuclear characteristics to future studies. Histopathological characterization of liver tumors is often challenging for hepatopathologists, with significant inter-observer variability frequently observed. Leveraging two large datasets of hematoxylin and eosin-stained digital slides, Liao et al. employed convolutional neural networks (CNNs) to distinguish HCC from adjacent normal tissues, achieving Areas Under the Curve (AUC) values exceeding 0.90 [108]. Similarly, Kiana et al. developed a tool to classify image patches as either HCC or CCA, attaining an accuracy of 0.88 on the validation set. Notably, their findings revealed that combining the model's predictions with pathologist input outperformed both the model and pathologist working independently. This emphasizes that AI tools are best utilized as augmentative aids rather than replacements for traditional histological diagnosis. However, they also highlighted that incorrect model predictions could adversely impact pathologists' final decisions, underscoring the importance of exercising caution when deploying AI for diagnostic automation [109]. The morphological characteristics of HCC significantly influence patient prognosis, prompting the development of various deep learning algorithms to enhance the prediction of HCC recurrence and survival using CT scans, MRI, or pathological images. Saillard et al. developed a model that processed HCC digital slides to predict patient survival following surgical resection with greater accuracy than conventional scoring systems incorporating clinical, biological, and PFs. Importantly, this model was validated across cases where slides were stained using different protocols, indicating its potential for broader generalizability across clinical centers [110].

Similarly, a recent study by Yamashita et al. [111] further confirmed the effectiveness of AI algorithms in predicting outcomes from digital histologic slides. Lu and Daigle [112] utilized three advanced convolutional neural networks (VGG 16, Inception v3, and ResNet50), pretrained on ImageNet, for feature extraction from HCC histopathological slides within the TCGA-LIHC cohort. They then identified survival-associated features through multivariable Cox regression analysis. While this study underscores the potential of histopathology-based outcome prediction, its conclusions are limited by the absence of adjustments for other prognostic factors and the lack of validation using an external cohort.

In addition, Saito et al. [113] applied traditional ML techniques to handcrafted features from whole slide images in a relatively small cohort of 158 HCC patients. They developed a combined model capable of predicting HCC recurrence post-resection with an accuracy of 89%. Validation of these promising findings in larger cohorts will be an essential next step. A Study by Haotian Liao [114], focuses on utilizing deep learning, specifically convolutional neural networks (CNNs), to improve the diagnosis and mutation prediction of HCC from histopathological images. The research addresses the challenges of pathological diagnosis and mutation prediction in HCC by employing CNN models on H&E-stained histopathological slides. Datasets from theTCGA and the Biobank of West China Hospital (WCH) were used, featuring both WSIs and tissue microarrays (TMAs). The CNN model achieved high accuracy in distinguishing HCC from normal tissues, with areas under the ROC curve surpassing 0.90. Additionally, the model predicted somatic mutations in several genes associated with HCC, achieving notably high predictive accuracy for key genes like CTNNB1 and TP53.

Another study [112] aims to utilize CNNs to extract relevant features from histopathological images to aid in the prognosis and survival prediction for patients with HCC. Three pre-trained CNN models, VGG 16, Inception V3, and ResNet 50, were used to analyze 421 tumor samples and 105 normal tissue samples from the TCGA database. These CNNs were employed for feature extraction, enabling classification and prognostic modeling. The CNN models effectively distinguished between cancerous and normal samples. Using a Cox proportional hazards model, significant correlations were found between extracted image features and overall survival, as well as disease-free survival, with concordance indices reflecting accurate survival predictions. The research demonstrates that pre-trained CNNs can effectively process histopathological images to extract prognostic markers, facilitating a deeper understanding of HCC progression and supporting clinical decision-making.

This study introduces a novel approach to interpreting AI-derived features in HCC through the development of a Practical Pathobiological Dictionary. To compensate for the limited availability of similar research, we defined key features with semantic relevance to tumor grading and employed nine feature selection models to identify the top 20 features most predictive of survival outcomes. However, several limitations should be noted. Despite the novelty of the proposed method, we are subject to errors due to limited external validation and a small number of samples to increase validity. In addition to trying to bridge computational and clinical data, the issue of decision-making transparency of AI models remains an ongoing challenge. While we aimed to define features using validated references, the small sample size and intra-tumoral heterogeneity may affect the generalizability and reliability of some feature interpretations. Importantly, the dictionary itself is still in the early stages, and further evaluation is needed to assess its consistency, clinical utility, and adaptability across different datasets and clinical scenarios. Additionally, our analysis focused primarily on general tissue and cellular features, without a detailed exploration of nuclear substructures or microenvironmental context. Future studies should incorporate advanced image analysis techniques, explore tumor–non-tumor distinctions, and further validate both the dictionary and the selected features to enhance clinical integration and interpretability of AI-driven models.

## 5. CONCLUSION

In this study, we developed a dictionary that allows features derived from image analysis of pathology slides to be mapped to the semantic features associated with the WHO Grading System for HCC. Researchers can use this dictionary to better understand the definitions related to pathologic semantic features. AI programmers can benefit from these definitions to learn how to handle numerical features in a way that makes them interpretable for healthcare professionals and suitable for clinical practice. Finally, to ensure that the definitions and the proposed connections to semantics are valid, we used statistical methods and feature selection algorithms. The key imaging features selected by the best-performing algorithm (accuracy: 0.80), including nuclear and cytoplasmic staining, are closely linked to cellular differentiation—an essential criterion explicitly emphasized in the WHO grading system. Additionally, when combined with imaging features, clinical variables such as TNM stage, age, gender, and race play a well-established role in determining and predicting patient outcomes. An important consideration is that, although this study was specifically designed for HCC, many of the identified features are shared across different diseases and their pathologies. As a result, this work can serve as a foundational framework for application in cancers of other organs as well.

**DATA AND CODE AVAILABILITY**. All codes and tables are publicly shared at:
*https://github.com/MohammadRSalmanpour/Practical-Pathobiological-Dictionary-Defining-Pathomics-Features/tree/main*

**ACKNOWLEDGMENTS.** This work was supported by the Mitacs Accelerate program grant number AWD-024298-IT33280. This work was supported by the Canadian Foundation for Innovation-John R. Evans Leaders Fund (CFI-JELF) program [Grand ID 42816]. We acknowledge the support of the Natural Sciences and Engineering Research Council of Canada (NSERC), [RGPIN-2023-03575] and Discovery Horizons Grant [DH-2025-00119]. Cette recherche a été financée par le Conseil de recherches en sciences naturelles et en génie du Canada (CRSNG), [RGPIN-2023-0357]. We also acknowledge the support from Virtual Collaboration (VirCollab, *www.vircollab.com*) Group as well as Technological Virtual Collaboration Corporation (TECVICO CORP.), Vancouver, Canada.

**CONFLICT OF INTEREST.** The authors have no relevant conflicts of interest to disclose.

*Salmanpour et al.*  *Dictionary Version LCP1.0*